%Paper: hep-th/9507087
%From: biru@theory.tifr.res.in (Bireswar Basu-Mallick)
%Date: Mon, 17 Jul 95 16:17:36 -2359

\magnification = \magstep1
\baselineskip = 24 true pt
\hsize = 17.3 true cm
\vsize = 22.5 true cm
\vskip .75 true cm
\centerline {{\bf Algebraic aspect and construction of Lax operators
in quantum integrable systems }}
\vskip  2.5 true cm
\centerline {
B Basu-Mallick$^{\dagger }$ and Anjan Kundu$^{\dagger \dagger}$  }
\vskip .25 true cm
\centerline {\it $^{\dagger  }$Tata Institute of Fundamental Research, }
\centerline {{ \it  Theoretical Physics Group, }}
\centerline {{ \it  Homi Bhabha Road, Bombay-400 005, India }}
\vskip .25 true cm
\centerline {{\it $^{\dagger \dagger}$Theory Group,}}
\centerline {{\it  Saha Institute of Nuclear Physics }}
\centerline {{\it Block AF, Sector 1, Bidhannagar Calcutta-700 064, India }}
\vskip 3.5  true cm
\noindent {\bf Abstract}

An algebraic construction more general and intimately connected
with that of Faddeev$^1$, along with its application for generating
different classes of quantum integrable models are summarised
to complement the recent results of ref. 1
( L.D. Faddeev, {\it Int. J. Mod. Phys. } {\bf A10},  1845 (1995) ).

\hfil \break \vfil \eject

A significant portion of  an excellent recent review
by Faddeev$^1$ is devoted to
an important result showing deep relation between the Yang--Baxter
equation and the quantum group. We are delighted to note that a scheme
for systematic construction of the Lax operators through Yang--Baxterisation
of the  Faddeev--Reshetikhin--Takhtajan (FRT) algebra
persued by us over the last
few years$^{2,3}$  has an intimate connection    with  this approach.
In  fact  our formulation gives a more general framework allowing construction
of a wide range of Lax operators of lattice models and recovers the
corresponding result of ref 1
related to $U_qsl(2)$
 as a particular case. Another
convincing demonstration of the usefulness of our approach is that, the
well known Lax operator of the lattice Liouville model$^4$ as well as
the recently discovered$^5$ nontrivial spectral
 parameter dependent variant of it
can be constructed directly and easily from our general form, instead of
going through the
involved limiting procedure and latice gauge transformation,
as is the standard practice$^{4,5}$.

We also consider the {\it twisted} generalisation of our Lax operator
as well as the $q \rightarrow 1$ limit of it and apply them for generating
different classes of integrable models.

Therefore we  hope that,  the present letter would
complement the relevant results of ref. 1.
                      and would
serve as a guide to understand and formulate the related
approach of Faddeev
in a more general form  and apply it for constructing wider class
of quantum integrable models.

                                \smallskip

For generating the class of  models associated with
the trigonometric  $R$-matrix of  spin-${1\over2}$  XXZ  chain:
$$
R_{trig}(\lambda) = \pmatrix{
\sin \alpha(\lambda+1) \ & \quad \ & \quad \ & \quad \cr
    \quad \ & \sin \alpha\lambda \ & \sin \alpha \ & \quad  \cr
     \quad \ & \sin \alpha  \  &
     \sin \alpha\lambda \ & \quad \cr
        \quad \  & \quad \ & \quad \  & \sin \alpha(\lambda+1)
        }.
\eqno (1) $$

we start with a general Lax operator of the form$^2$
 $$ L(\lambda ) ~=~  \pmatrix {   { {1\over \xi } \tau_1^+  +
 \xi \tau_1^- } & {\tau_{21}^{~} }  \cr
    { \tau_{12}^{} }  &   {  {1\over \xi } \tau_2^+  + \xi \tau_2^-  }     }
 ~ . \eqno (2) $$
Here $ \xi = e^{ - i  \alpha \lambda  } $ is the spectral parameter
and the  operator valued elements in $L(\lambda )$ satisfy the algebra
$$ \left [ \tau_{12}, \tau_{21} \right ] ~=~
-2i \sin {\alpha} \left (  \tau_1^+ \tau_2^-  -  \tau_1^- \tau_2^+ \right ) ~,~
\tau_i^{\pm }\tau_{ij} ~=~e^{\pm i\alpha } \tau_{ij} \tau_i^{\pm }~,~
\tau_i^{\pm }\tau_{ji} ~=~e^{\mp i\alpha } \tau_{ji} \tau_i^{\pm }~,~
  \eqno  (3) $$
( $i,j ~\in ~ [1,2] $ ) with all  $~\tau_i^{\pm} ~$
commuting  among  themselves.
The Hopf algebra structure of (3)
including the coproduct, unity,  antipode etc.  can  be shown to exist in the
usual way.
It is also worthnoting that the above  quadratic algebra
is an extension of the trigonometric Sklyanin algebra (TSA)$^{6}$
  and reduces to it
in the particular symmetric case $~\tau_2^- ~=~ - \tau_1^+ ~,
{}~~\tau_2^+ ~=~ - \tau_1^- ~$.
Moreover,  in analogy with the well known relation$^7$ between TSA and
$U_qsl(2)$,
 one can  express  the elements  of extended TSA (3)
  as$^{3,8,9}$
$$
\tau_1^\pm =   \left( \tau_2^\pm \right)^{-1}  =
  q^{ \pm  S_3 }~,~~\tau_{12} = - ( q-q^{-1} ) S_+ ~,~~
 \tau_{21} =  ( q-q^{-1} ) S_- ~,  \eqno (4)
$$
where
 $ q = e^{i\alpha } $ and
 $S_3 , ~ S_{\pm }$ are the generators of the  quantised
algebra $U_qsl(2)$. For the above realisation of its operator elements,
 our Lax operator (2)  coincides with
the Lax operator of $q$-deformed  XXX-model given by eqn. (126)
in ref. 1. Furthermore, by using algebra (3),  it is easy to check that
the $L(\lambda )$ operator (2) satisfies the quantum Yang--Baxter
equation (QYBE)
$$ R(\lambda - \mu)~ L_1(\lambda)~ L_2(\mu ) ~=~
L_2(\mu )~ L_1(\lambda)~ R(\lambda - \mu) ;
{}~~ L_1 \equiv  L \otimes {\bf 1} ,~ L_2 \equiv {\bf 1} \otimes L ~
\eqno (5) $$
for   the trigonometric $R$-matrix (1) and thus may be considered
as the Lax operator of some abstract integrable model, concrete
realisations of which should yield different physical models.

To extract the spectral parameterless limit of $L(\lambda )$
 and the  corresponding $R(\lambda )$ given by (2) and (1),
 it is helpful to make
 a `gauge transformation' on them ( see eqns. (2.5)-(2.7)
of ref.2  or,
 eqns. (131),  (132) of ref. 1 ), which
 allows us to write them
   as
$$
    R(\lambda ) ~=~ {1\over \xi } R^+ ~-~ \xi R^-  ~,~~~
  L(\lambda  )  ~=~  {1\over  \xi  }  L^{(+)} ~+~\xi L^{(-)} ~, \eqno
(6)
$$
where
$$
L^{(+)} ~=~ \pmatrix {  {\tau_1^+} & {\tau_{21}} \cr {0} & {\tau_2^+} } ~,~~
L^{(-)} ~=~ \pmatrix {  {\tau_1^-} & {0} \cr {\tau_{12}} & {\tau_2^-} } ~.
\eqno (7)
$$
Again,  for
 realisation (4),
the above $L^{({\pm })}$-matrices  coincide with
their counterparts  given through  eqns. (139) and (140)
of ref. 1.

Subsequently,
we may  insert, as shown in ref. 2, the  `gauge transformed'
 $L(\lambda )$ and $R(\lambda )$
matrices (6) in the QYBE (5) and compare  the  coefficients
of various powers of spectral parameters from its both sides and arrive
at a set of seven relations
$$
 R^{\pm}~L^{(\pm )}_1~ L^{(\pm)}_2 ~=~L^{(\pm)}_2 ~L^{(\pm)}_1~R^{\pm} ~,~~
 R^{\pm}~L^{(\pm)}_1~ L^{(\mp)}_2 ~=~L^{(\mp)}_2 ~L^{(\pm)}_1~R^{\pm}~,~
 \eqno (8)
 $$
and
$$
   R^{+}L^{(-)}_1 L^{(+)}_2 ~-~L^{(+)}_2 L^{(-)}_1 R^+ ~=~
   R^{-}L^{(+)}_1 L^{(-)}_2 ~-~L^{(-)}_2 L^{(+)}_1R^- ~. \eqno (9)
$$
which are
evidently similar to  eqns. (144)-(147)
of  ref. 1.
By using  once again realisation    (4), it is easy to see that  the
  relations (8) are essentially  same as the
  well known FRT relations for the
quantised algebra  $U_qsl(2).$ Moreover, the remaining eqn. (9) can also
be reduced to these FRT relations with the help of
crucial Hecke condition
$$   R^+  ~-~R^-  ~=~c ~{\cal P},\eqno (10) $$
( ${\cal P}$ being the permutation operator )
 satisfied by $R^\pm $ matrices$^{2,1}$.
Thus, as  we have  shown in ref. 2,  all
spectral parameter independent   FRT relations corresponding to
   the quantised algebra can in fact be derived from a
single, but spectral parameter dependent  relation
 QYBE.

We shall stress again that  by using   different realisations
 of extended TSA (3) in
 the general Lax operator (2),
one can easily construct  the representative Lax operators of a
large set  of quantum integrable models, all of them associated
with the same trigonometric $R$-matrix (1).
Examples of such models are the lattice sine-Gordon
model (for the symmetric
reduction (4)), lattice Liouville model, an ultralocal derivative
nonlinear Schr\"odinger  (NLS) equation, massive Thirring model
etc.
including some new ones$^3$. Moreover,
 reverting the above outlined procedure,
 it is    possible to Yang--Baxterise the FRT algebra
 in a straightforward way
 and
 construct  other solutions of QYBE   including
 those with
higher dimensional Lax operators$^{2,10}$.

The following is the reason, why
it is often easier to  construct different types of
 Lax operators of quantum integrable
models from
    $L(\lambda )$ operator  (2), in comparison with the
 Lax operator of $q$-deformed XXX-model given by eqn. (126)
in ref. 1. The extended TSA (3) yields a
  realisation like (4)  through the generators of
the quantised algebra  $U_qsl(2),$
 only  when all $\tau_i^\pm $ are invertible nonsingular operators.
However, for singular values of some $\tau_i^\pm $,
 there  exist  other realisations of extended TSA
which cannot be connected with  $U_qsl(2)$ algebra, without
making  any complicated limiting transition.
 To illustrate this important point, we may
consider the   example of  lattice Liouville model.
Recall  that
for constructing
the associated well known  Lax operator$^4$  as well as the Lax
operator
with  nontrivial spectral parameter dependence discovered
very recently$^5$,  one starts usually  with
 an infinite dimensional representation of $U_qsl(2)$ quantum
algebra, which
maps the $q$-deformed $XXX$-spin Lax operator to
 the lattice sine-Gordon Lax  operator. Subsequently one
has to perform a rescaling of the field  and then
take the massless limit  for  obtaining  the standard Lax operator$^4$.
The  construction of the recent  nontrivial variant  of it
is even more involved. Along with the above rescaling of the field and
the massless limiting procedure, it  requires also a
renormalisation of the  spectral parameter in addition to a
lattice gauge transformation on the sine-Gordon Lax operator$^5$.
It is
  however interesting to  observe that,
one can use   simple realisations
of extended TSA  (3) to  get in a straightforward way,
both the standard and the nontrivial
Lax operators of the lattice  Liouville model
from our $L(\lambda )$ operator (2),
without involving  any
rescaling,
 gauge transformation or  limiting procedure.
 It is easy to check that the realisation of (3) through
 canonical operators
 $ \Phi, ~\Pi $ ( with
 $ [\Phi , \Pi ] = i \alpha $ ) as
$$
 \tau_1^+ =\tau_2^- =
 -i\, e^{-i \Phi }~,~~ \tau_2^+   ~=~\tau_1^- ~=0,~
 \tau_{12} ~=~e^{i\Pi }~h^{{1\over2}}(\Phi ) ~~,~~
\tau_{21} ~=~h^{{1\over2}}(\Phi ) ~e^{-i \Pi } , \eqno (11)
$$
where $~ h(\Phi ) = 1 - e^{ -2i \Phi + i \alpha }, $
is consistent with algebra (3) and  gives directly from (2) the well known
Lax operator$^4$ of the lattice Liouville model. Moreover, in an analogous way
through another simple and similar realisation
$$
 \tau_1^+ =\tau_2^- = ( \tau_1^-  )^{-1} ~=~
 -i\, e^{-i \Phi }~,~~\tau_2^+ =0~,~~\tau_{12} ~=~e^{i\Pi }~,~~
\tau_{21} ~=~h(\Phi ) ~e^{-i \Pi } , \eqno (12)
$$
we  obtain readily from  (2)
 the nontrivial spectral parameter dependent
 Liouville Lax operator
of ref.5 without going through any other intermediate steps.

Apart from the construction related to the quantum $R$-matrix (1)
described above, there exist  also other interesting
possibilities to  cover wider range of integrable models.
 If we consider a {\it twisting} transformation of the $R$-matrix (1):
 $$
R\rightarrow R_\theta=B(\theta)RB(\theta), \qquad B(\theta)
                  = e^{i{\theta }(\sigma_3\otimes 1-1\otimes \sigma_3)},
\eqno (13)
$$
 the $L(\lambda)$ operator as a corresponding solution
of the QYBE (5) may again  be given in the form (2),
where the generators
$\{ \tau \}$ now satisfy a $\theta$-deformed extension
of the quadratic algebra (3)
$$
 e^{i\theta}~ \tau_{12} \tau_{21}~ -~ e^{-i\theta}~ \tau_{21} \tau_{12}
  ~=~
 -2i \sin \alpha \left (  \tau_1^+ \tau_2^-  -  \tau_1^- \tau_2^+ \right )
 ~, $$ $$
\tau_i^{\pm }\tau_{ij} ~=~e^{i(\pm \alpha+\theta)}
 \tau_{ij} \tau_i^{\pm }~,~\ \
\tau_i^{\pm }\tau_{ji} ~=~e^{i(\mp \alpha-\theta)} \tau_{ji} \tau_i^{\pm }~.~
\eqno (14)$$
Note that the decomposition (6) and the relations (8-10) also hold
good in this case.
 Symmetric reduction like (4) relates this algebra to the two-parameter
 algebra $U_{q,p}gl(2)$
  and may generate $\theta$-deformed lattice sine-Gordon
and Liouville model.
However, other possible realisations yield from (2) the Lax operators
of the well known Ablowitz--Ladik model and a family of
 discrete-time models related to the   relativistic quantum
Toda chain$^{3,11}$.

Likewise, using slightly different type of twisting one
can also  Yang-Baxterise  the  FRT algebra  related to coloured
braid group representation and generate
 integrable Lax operators associated  with non-additive type spectral
parameter dependent quantum  $R$-matrix$^9$.
A general formalism in this direction starting
from an universal $R$-matrix of reductive Lie algebras is  also presented
in a recent work$^{10}$.

\smallskip

It is  remarkable that most of  these structures survives nicely  in the
$q\rightarrow 1$ or $ \alpha \rightarrow 0$
limit, when the trigonometric $R$-matrix reduces to its rational form
$$
R_{rat}(\lambda) = \pmatrix{
 \alpha(\lambda+1) \ & \quad \ & \quad \ & \quad \cr
    \quad \ &  \alpha\lambda \ &  \alpha \ & \quad  \cr
     \quad \ &  \alpha  \  &
    \alpha\lambda \ & \quad \cr
        \quad \  & \quad \ & \quad \  &  \alpha(\lambda+1)
        },
\eqno (15) $$
and allows us to generate another class of integrable
quantum models associated with this rational $R$-matrix solution.
At this undeformed limit  the $L(\lambda)$ operator (2) also
undergoes a smooth transition to the form$^3$
$$ L(\lambda )
 = \pmatrix {  {K_1^0 + i  { \lambda \over \eta } K_1^1 }  &   {K_{21}}
\cr
  {K_{12}} & {K_2^0 + i  { \lambda \over \eta } K_2^1 } \cr  } , \eqno (16) $$
where {\bf K } operators now satisfy  another quadratic
algebra being the $q \rightarrow 1$ limit of (3):
$$ \eqalign {  [K_{12} , K_{21} ] &
 =  ( K_1^0 K_2^1 - K_1^1 K_2^0 ) , ~~~[ K_1^0, K_2^0 ] = 0
\cr
 [K_1^0, K_{12}]  =   K_{12} K_1^1 ,&~ [K_1^0, K_{21}]  = -  K_{21}K_1^1,
{}~~ [K_2^0, K_{12}] = -   K_{12} K_2^1 ,~ [K_2^0, K_{21}] =  K_{21} K_2^1
,  }  \eqno (17) $$
with $ K_1^1, K_2^1 $ serving as Casimir operators.
A particular symmetric  reduction $K_1^1 =K_2^1  =  1
$
and $ K_1^0  =  -K_2^0  $  yields  the  standard  $sl(2)$  algebra,
which for proper realisation gives $XXX$ spin-${1\over 2}$ model
and the lattice NLS model. However using the  freedom of   other
realisations of (17), from    $L$-operator (16) one may obtain the
nonrelativistic Toda chain and  a simple version of lattice NLS$^{12}$
and through twisting transformation
as before the  $\theta$-deformed  models.

The classical aspect of our approach and its application are discussed in
ref. 13.

\vfil \eject

\noindent {\bf References }

\item {1.}  L.D. Faddeev, {\it Int. J. Mod. Phys. } {\bf A10},  1845 (1995).
\item {2.} B. Basu-Mallick and A. Kundu, {\it J. Phys. } {\bf 25}, 4147
(1992).
\item {3.}
A. Kundu  and B. Basu-Mallick, {\it Mod. Phys. Lett. } {\bf A7},  61 (1992);
\item {}  B. Basu-Mallick  and A. Kundu,
 {\it Phys. Lett. } {\bf  B287}, 149 (1992);
\item {} A. Kundu  and B. Basu-Mallick, in { Proc. Of Int. Conf. Needs' 91,
Gallipoli, Italy }
( Ed. M. Boiti, L. Martina and F. Pompinelli, World Sc., 1992 ) p. 357;
\item {} A.Kundu and B.Basu-Mallick, {\it J.Math. Phys. }
{\bf 34 }, 1052  (1993).
 \item {4.} L. D. Faddeev and L. A. Takhtajan,
in {\it Integrable Quantum Field Theories, Lecture notes in Physics , }
eds. H. J. de Vega et al. (Springer Verlag, Berlin,1986) Vol. 246, p. 166.
\item {5.} L.D. Faddeev and O. Tirkkonen,  {\it Connections of
the Liouville model and XXZ spin chain }, HU-TFT-95-15, hep-th/9506023
(1995).
\item {6.} E.K. Sklyanin, {\it Funk. Anal. Pril } {\bf 16}, 27 (1982).
\item {7.}
 H.  Saleur  and J. B.  Zuber,   {\it  Integrable lattice models
and quantum groups, }  Saclay preprint, SPhT/90-071 (1990).
\item {8.} C. K. Zachos, {\it private communication }
\item {9.} A. Kundu  and B. Basu-Mallick, {\it J. Phys. } {\bf A27 },
3091  (1994);
\item {} B. Basu-Mallick, {\it Mod. Phys. Lett. } {\bf A9}, 2733  (1994).
\item {10.}
Anjan Kundu
and P.Truini {\it Universal $ R$-matrix of reductive Lie algebras }
to appear in {\it  J. Phys.A } (1995)
\item {11.}
  Anjan Kundu,  {\it Phys.Lett.  } {\bf A190}, 73 (1994)
\item {12.}
       Anjan Kundu and  O. Ragnisco,  {\it J.Phys.  }{\bf A27}, 6335 (1994)
\item{13.}
 Anjan Kundu,
 {\it  Teor. Mat. Fiz.}
 {\bf 99}, 428 (1994)

\end